\documentclass[a4paper,fleqn]{cas-dc}

\usepackage[numbers]{natbib}

\usepackage{amsmath}

\DeclareMathOperator{\rect}{rect}
\DeclareMathOperator{\Circ}{Circ}
\def\tsc#1{\csdef{#1}{\textsc{\lowercase{#1}}\xspace}}
\tsc{WGM}
\tsc{QE}
\tsc{EP}
\tsc{PMS}
\tsc{BEC}
\tsc{DE}

\begin{document}
\sloppy
\let\WriteBookmarks\relax
\def\floatpagepagefraction{1}
\def\textpagefraction{.001}
\shorttitle{Photonic crystal fiber for lensless microscopy}
\shortauthors{S Kumar et~al.}

\title [mode = title]{Photonic crystal fiber for high resolution lensless in-line holographic microscopy}                      



\author[1]{Sanjeev Kumar}[orcid=0000-0002-9943-6343]
\cormark[1]
\fnmark[1]
\ead{sanjeevsmst@iitkgp.ac.in}


\address[1]{School of Medical Science and Technology, Indian Institute of Technology, Kharagpur, 721302, India}

\author[1]{Manjunatha Mahadevappa}[]

\author[2]{Pranab K. Dutta}[]

\address[2]{Department of Electrical Engineering, Indian Institute of Technology, Kharagpur, 721302, India}

\cortext[cor1]{Corresponding author}


\begin{abstract}
We propose to use high numerical aperture single mode optical fibers like photonic crystal fiber for lensless in-line holographic microscopy. Highly divergent beam helps to overcome the spatial sampling limitation of the image sensor. In this paper, a submicron lateral resolution has been demonstrated, with an imaging sensor of pixel pitch 1.12 $\mu$m and a photonic crystal fiber of mode field diameter 1.8 $\mu$m. In earlier methods of single-shot lensless imaging, submicron resolution has been obtained at very small working distance and field of view. The proposed method improves the resolution without compromising the working distance. A working distance of (but not limited to) $\sim1.7$ mm with a field of View $\sim1.4$ mm has been demonstrated.
\end{abstract}




\maketitle

\section{Introduction}
\subsection{Lensless in-line holographic microscopy}
Lensless in-line holographic microscopy based on a coherent point source was invented by Dennis Gabor in 1948 to overcome the effects of aberrations in both the electromagnetic and optical lenses (see figure 1 for imaging setup) \cite{gabor1948new}. In Gabor in-line holography:
\begin{align}
&\text{Recorded Intensity:} \ I \approx U_o U_r^* +  U_o^* U_r \\
&\text{Reconstructed Field:} \ U_p I  \approx U_p U_o U_r^* +  U_p U_o^* U_r
\end{align}
where  $\ U_o$ denotes the scattered object wave-field, $\ U_r:$ unscattered reference wave-field, $\ U_p:$ reconstruction wave-field, all at the observation plane and $\ *$ denotes the complex conjugates.

The method has been extensively investigated with low numerical aperture (NA) optical sources like pinholes with diameter $\sim$100-200 micrometers and low NA single mode optical fibers for large field of view (FOV) and telemedicine applications. Limitations of this method include degradation of the reconstucted image by the twin image artifact, low resolution due to limited pixel pitch ($\sim 1 \mu$m) and small working distance when partially coherent light is used. Twin image artifact is suppressed iteratively using phase retrieval algorithms \cite{fienup1982phase,latychevskaia2007solution}. Resolution is improved using various super-resolution methods based on multiple shot imaging like ptychography \cite{maiden2011superresolution} and fourier ptychography (synthetic aperture imaging) \cite{luo2015synthetic}. These methods give significant improvement in resolution but increase both image acquisition and image reconstruction time. Reduced coherence length in case of partially coherent sources imposes a working distance limit (see \cite{ozcan2016lensless}). In most of the reported works, object to sensor distance $z_2$ of less than a millimeter has been demonstrated. Long working distance (in the range of few millimeters) is necessary for variety of imaging applications. Examples include imaging of cells or microbubbles flowing in capillaries or microfluidic channels and cells or tissue cultured in vials.

To overcome the low resolution in single-shot imaging, high NA point sources can be used. A strongly diverging coherent beam can be used to modulate the spatial frequency spectrum of scattered object wave during holographic recording. Some groups have utilized sources coupled to micron and submicron diameter apertures as high NA point sources \cite{jericho2011point,takaki1999fast}. Kanka et al has shown an imaging NA of 0.8 using a pinhole aperture of diameter 650 nm with their tile superposition algorithm \cite{kanka2009reconstruction,kanka2011high}. This method has a limitation that the illuminated region of the object should be very small as compared to the size of image sensor. Apart from FOV, it also limits the working distance i.e. limits source to object distance $z_1$ (see figure 1).

Here we propose an alternative, photonic crystal fibers (PCF) can be used as a high NA single mode coherent light source for digital in-line holographic microscopy (DIHM). It helps to improve the resolution for a large FOV and working distance.

The organization of this paper is as follows: first we discuss the principle of imaging and magnification, when a gaussian beam is used for lensless in-line holography. Then, we discuss the method of recording holograms using a photonic crystal fiber followed by the method of holographic reconstruction. We demonstrate the improvements in resolution and working distance using this principle. Then we present a theoretical discussion about the resolution and working distance which can be expected in lensless holography with a gaussian beam.

\subsection{Principle of magnification}
The principle of imaging and magnification which we have exploited in this paper has been discussed next.
A fundamental gaussian mode from an optical fiber gives a spherical phase variation in the far field region. 
\begin{equation}
\phi(x,y) = \exp( - i k z ) \exp( \frac{- i k \left( x^2 +y^2 \right)}{2  z }) 
\end{equation}
where $\ \phi $ denotes the phase of the gaussian field, $\ k $ is the wave vector, $\ x $ and $\ y $ are the transverse coordinates and $\ z $ is the axial distance from source to observation plane.

In equation 2, if the reconstruction wave $ U_p$ is a plane wave, the recreated field $ U$ contains an optically magnified copy of the object. The reason has been discussed next.
\begin{equation}
\text{Since $U_p=1$; Reconstructed Field:} \  U  \approx  U_o U_r^* +   U_o^* U_r
\end{equation}
The phase parts of the reference beam  $ \phi \left(U_r \right)$ and its conjugate $ \phi \left(U_r^* \right)$ in equation 4 act as lens functions.

Lens function after paraxial approximations \cite{goodman2005introduction}:
\begin{equation}
L(x,y)  \approx \exp( \frac{\pm i k \left( x^2 +y^2 \right)}{2  f }) 
\end{equation}
where f is the focal length.
\begin{equation}
\text{Phase of $U_r^*$ and $U_r$:}\ \phi(x,y)  \approx \exp( \frac{\pm i k \left( x^2 +y^2 \right)}{2  z }) 
\end{equation}
where z is source to sensor distance. This gives:
\begin{equation}
U  \approx  U_o L_{converging} +   U_o^*  L_{diverging}
\end{equation}
\begin{figure}
\centering
\includegraphics[width=7cm]{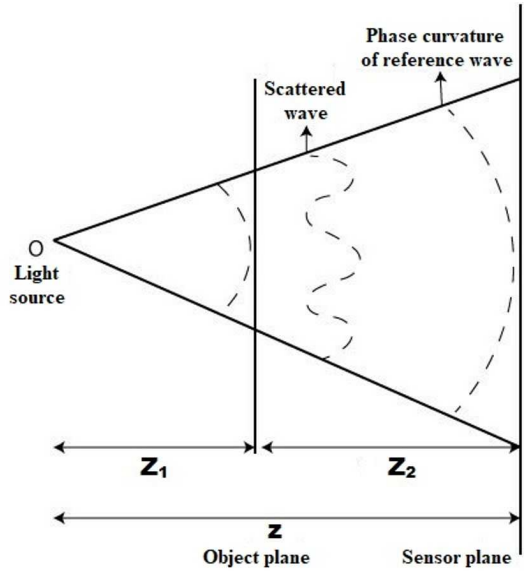}
\caption{Imaging geometry for lensless in-line holographic microscopy with a gaussian beam.}
\label{fig:false-color}
\end{figure}
\begin{figure}
\centering
\includegraphics[width=7cm]{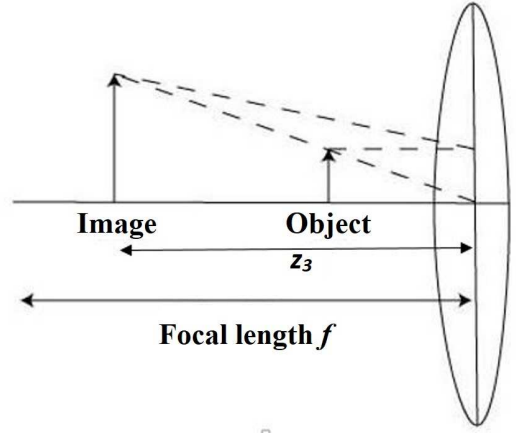}
\caption{Magnified virtual image in an imaging system.}
\label{fig:false-color}
\end{figure}
$ \phi \left(U_r^* \right) = L_{converging}$ acts like a convex lens and optically magnifies the image of the object. In other words, it modulates the spectrum of spatial frequencies carried by the object wave. Because of larger NA of PCF and hence larger spot size of beam, the phase curvature of the reference wave can be easily exploited for incorporating magnification during holographic recording. This optically magnified virtual image can be focused computationally using wavefront propagation. 

Note that the various groups have multiplied the hologram with a spherical reconstruction wave which is an approximation of reference wave($U_p \approx U_r$) to obtain original object wave $U_o$. In that case the final image will be obtained with unit magnification and resolution will again be limited by the pixel pitch. Then an algorithm like tile-superposition algorithm \cite{kanka2009reconstruction} or a planar screen to spherical surface coordinate transformation \cite{takaki1999fast} is required to calculate the high resolution image. 

\subsection{Digital focusing}
Various wavefront propagation methods have been investigated by many authors and their advantages and disadvantages have been established \cite{matsushima2009band,kozacki2012computation,kanka2009reconstruction}. We have used angular spectrum method for primary reconstruction. It is based on the Rayleigh-Sommerfeld integral and is acceptable in microscopy because of its validity in non-paraxial regime. It involves two FFT operations and a multiplication step (in Fourier domain) with the following coherent transfer function.
\begin{equation}
H(f_x,f_y)= \exp(- i \pi \lambda z_3 \left( f_x^2 +f_y^2 \right) )
\end{equation}
where $\ \lambda $ is the wavelength, $z_3$ is the distance of the image from the sensor, $\ f_x $ and $\ f_y $ are the spatial frequencies.

To improve the contrast of the reconstruction, we have used maximum-likelihood blind deconvolution\cite{markham1999parametric} on the amplitude of the object obtained by angular spectrum method. It is a statistical method of simultaneously estimating both point spread function and better approximation of reconstructed image. 
\begin{figure}
\centering
\includegraphics[width=8.25cm]{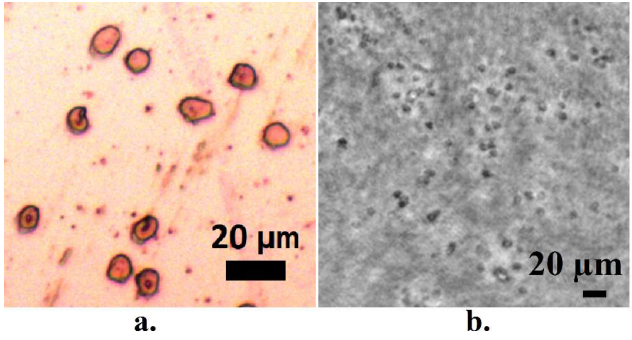}
\caption{(a) Red blood cells (diameter $\sim$5-7 microns) image captured in bright field microscope with 20x objective lens. (b) Lensless microscopy of same sample with a low NA optical fiber at magnification $\sim$1, shows a low resolution reconstruction.}
\label{fig:false-color}
\end{figure}
\section{Methods}
\subsection{Imaging experiments}
We have used a photonic crystal fiber (PCF) with the pure silica core as a high NA ($0.38 \pm 0.05$) coherent light source. The mode field diameter of this single mode optical fiber is $1.8\ \pm \ 0.3\ \mu m$. A pigtailed monochromatic laser source of $675\ nm $  wavelength and $2.5\ mW$ power has been connected to PCF fiber using FC/PC connector. An eight mega pixel image sensor with pixel pitch $1.12\ \mu m\ \times 1.12\ \mu m$ and physical size $3.68\ mm$ $\times$ $2.76\ mm$ has been fixed at a distance of about 2 to $ 3.4\ mm$ from source position (see figure 1). Object mounted on a micrometer stage is kept in the beam path. Different samples like coating stripped single mode optical fibers with core diameter 10 microns and blood slides containing RBCs (diameter around 5-7 microns) have been imaged. By changing the position of object, diffraction patterns at different magnifications ($M=\frac{z}{z_1}$) have been captured. Field of view is equal to the ratio of area of reference gaussian spot and magnification.

\subsection{Reconstruction}
Holographic reconstruction has been performed by multiplying hologram by a plane wave $U_p=1$. A virtual image on the same side of sensor as object but at a larger distance is obtained (see figure 2). Wavefront propagation is obtained by angular spectrum method using inverse of optical transfer function shown in equation 8. The focusing of reconstructed image is achieved iteratively by changing $z_3$ in equation 8 and assesed numerically by calculating the variance of gradient. The initial approximation of $z_3$ is magnification times $z_2$. i.e. $\frac{ z}{z_1} z_2$. 

Maximum-likelihood blind deconvolution is applied on the images obtained in previous step. For deconvolving image in figure 4(f), a kernel of all ones of size $ 5 \times 5$ has been used as initial estimate of point spread function. Result after 10 iterations has been shown in figure 4(g).

Twin image suppression step is optional depending on the extent of degradation and availablity of computational time. This artifact degrades image at unit magnification to a larger extent (compare figure 4(b) and 4(c)). When magnification = 2 or more, the apparent distance between the two copies of object is significantly large. Defocused twin image is spread over a larger area and hence less pronounced in the reconstruction.
 If necessary, twin image suppression has been done in this paper by phase retrieval algorithms described in reference \cite{fienup1982phase,latychevskaia2007solution} (figure 4(d) shows result after phase retrieval). 
\begin{figure*}
\centering
\fbox{\includegraphics[width=17.25cm]{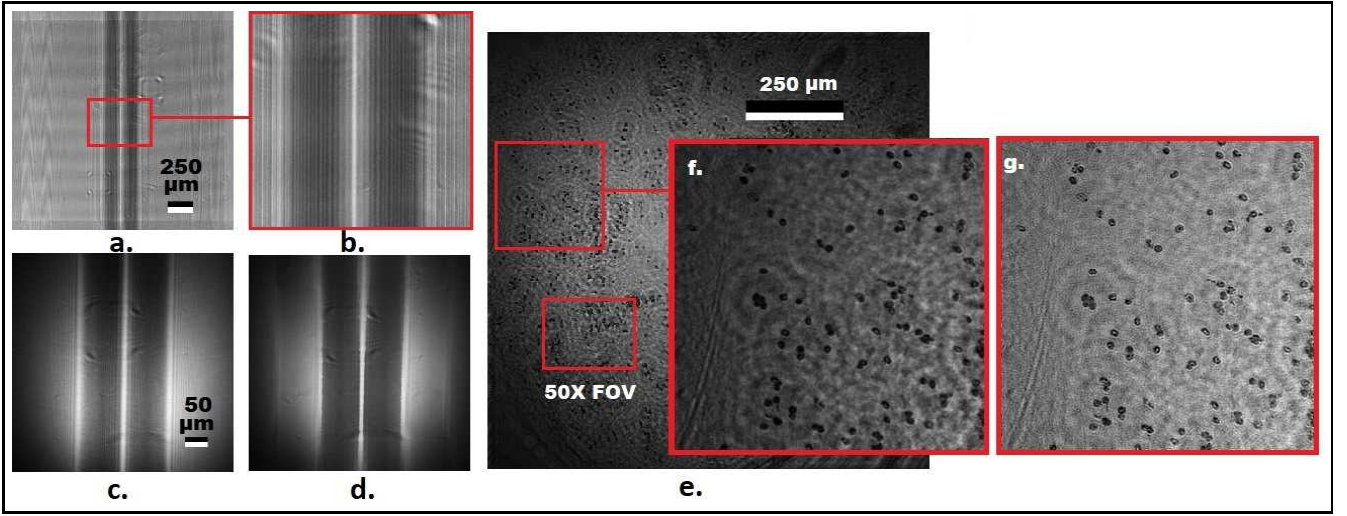}}
\caption{Lensless microscopy with photonic crystal fiber: (a) sample is an optical fiber with core diameter of 10 microns at magnification $\sim$1. (b) Smaller region of same image showing twin image artifact. (c) same sample at optical magnification $\sim$3. (d) Result after phase retrieval applied in c. (e-f) sample is RBCs at optical magnification $\sim$2. (g) Result after maximum-likelihood blind deconvolution applied on f.}

\label{fig:false-color}
\end{figure*}
\begin{figure}
\centering
\includegraphics[width=8.25cm]{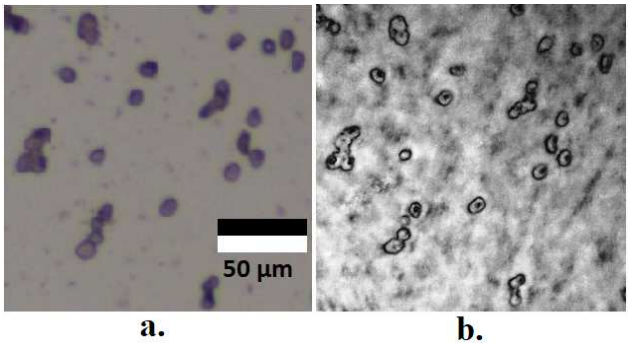}
\caption{(a)RBC image captured in bright field microscope with 10x objective lens. (b) Lensless microscopy with photonic crystal fiber, at optical magnification $\sim$3.}
\label{fig:false-color}
\end{figure}
\begin{table*}
\centering
\caption{\bf A comparison with the achromat flat field objectives of microscope \cite{Nikon}.}
\begin{tabular}{cccc}
\hline
\rule{0pt}{1\normalbaselineskip}
Objective&Resolution (in $\mu$m) &FOV &Working Distance\\
/Lensless&&Diameter (in mm)& (in mm)\\[2pt]
\hline
\rule{0pt}{1\normalbaselineskip}
10 x & 1.6 & 2.2 & 7 \\[2pt]
20 x & 1 & 1.1 & 3.9 \\[2pt]
40 x & 0.6 & 0.55 & .65 \\[2pt]
\hline
\rule{0pt}{1\normalbaselineskip}
Lensless & $\sim$1 & 1.4 & 1.7 (source to object)  \\
(Mag = $\sim$2) &&&1.7 (object to sensor)\\[2pt]
\hline
\rule{0pt}{1\normalbaselineskip}
\end{tabular}
  \label{tab:shape-functions}
\end{table*}
\section{Results and Discussion}
Figure 3 demonstrates lensless in-line holographic microscopy with a low NA light source. The magnification can be approximated around 1. The rbc features are not resolved because of low resolution. Figure 4e-g and 5 demonstrates lensless in-line holographic microscopy with a photonic crystal fiber at magnifications around 2 and 3. The rbc features in these images can be clearly observed because of the improved resolution. Figure 4a and c demonstrates the reduction in twin image artifact by introducing magnification. Table 1 shows the comparison of the resolution, field of view and working distance with a lens based microscope.

\subsection{Two-point resolution in gaussian beam }
The two point resolution of a coherent optical imaging system depends on the wavelength $ \lambda $ , numerical aperture $ n \sin \sigma $ and the phase difference of the two points being observed \cite{born2013principles,kohler1981abbe,horstmeyer2016standardizing}. In a circular pupil based system, when the sample is illuminated by a plane coherent beam at an angle, the following equation defines the ability to resolve two points:
\begin{equation}
\Delta = \frac{k \lambda}{n \sin \sigma}
\end{equation}
$k$ ($=$0.5 to 1) depends upon the phase difference of the two points under observation. The best resolution that can be obtained is for $k=0.5$ when the phase difference is odd multiple of $ \pi /2$ .

Here in gaussian beam holography, the illumination of object is gaussian function, the phase difference of two points under observation will depend on their transverse location on the object plane. Thus we will get a spatially varying two-point resolution. This can be seen mathematically in the following equations (shown for one-dimensional case for simplicity).

Consider two unit impulses at a distance $ \Delta= 2 x_1 $
\begin{equation}
 T(x)= \delta(x - x_1) +\delta(x+x_1)
\end{equation}
A Gaussian illumination $ G(x)$ at the plane of $\  T(x) $ will result in an optical field:
\begin{align}
T(x_1) G(x_1) = B \exp( \frac{- i k x_1^2}{2 z_1} ) \\ T(x_2) G(x_2) =  B \exp( \frac{+i k x_1^2}{2 z_1}
)
\\ T(x \neq x_1,x_2) G( x \neq x_1,x_2) = 0 
\end{align}
For simplicity, amplitude B has been approximated equal at these two points under observation. 

\subsection{Gaussian pupil}
The output of imaging system will be geometrically magnified image convolved with Point image function PSF.
\begin{align}
U_i(\xi) = T(x/m) G(x/m) \otimes PSF(\xi) \\
\text{Observed Intensity:}\ I_i(\xi) = U_i(\xi) U_i^*(\xi)
\end{align}
These equations show the dependence of two point resolution on both $ G(x)$ and $PSF$. 

 Point spread function can be obtained theoretically using Rayleigh-Sommerfeld integral \cite{goodman2005introduction} (contribution of twin image and other components of holography have been neglected). 
\begin{multline}
PSF(\xi , \eta)= \\
\iint_{-w}^{w} h_f(x,y,z_2) U_r^*(x,y,z) h_i(x- \xi,y- \eta,z_3) dx dy 
\end{multline}
where w is the pupil radius.
\begin{equation}
h_f(x,y,z)=\frac{1}{j \lambda} \frac{\exp(jkr)}{r} cos \ \theta; \
r= \sqrt{x^2+y^2+z^2}
\end{equation}
\begin{equation}
h_i(x,y,z)=IFT \Bigg[ \frac{1}{FT[h_f(x,y,z)]} \Bigg]
\end{equation}
Amplitude of the reference beam $U_r^*(x,y)$ acts as a Gaussian pupil. 
\begin{equation}
\text{Amplitude of $U_r$: Amp} (x,y)  \approx \exp( \frac{-   \left( x^2 +y^2 \right)}{a^2(z) }) 
\end{equation}
where $ a $ is the spot size and depends on $ z $. $a$ is generally restricted to radius where normalized intensity is around $\sim$ 0.14. 

As evident from the above discussion, due to gaussian illumination at object plane and gaussian nature of pupil, Abbe two point resolution becomes a complicated criterion for defining resolution here. Instead the amplitude transfer function cut-off frequency will be a better method for defining resolution here \cite{horstmeyer2016standardizing}.  This is basically dependent on the pupil function \cite{goodman2005introduction}. 
\subsection{Cut-off frequency}
Pupil function when gaussian spot size is larger than image sensor:
\begin{equation}
P(x,y) = \rect \left( \frac{x}{2 w_x} \right) \rect \left( \frac{y}{2 w_y} \right)
\end{equation}
Corresponding amplitude transfer function: 
\begin{equation}
H_p(f_x,f_y) = \rect \left( \frac{\lambda z_2 f_x}{2 w_x} \right) \rect \left( \frac{\lambda z_2 f_y}{2 w_y} \right)
\end{equation}
Pupil function when Gaussian spot size is smaller than image sensor:
\begin{equation}
P(x,y) = \Circ \left( \frac{\sqrt{x^2 +y^2}}{a} \right) 
\end{equation}
Corresponding Amplitude Transfer Function:
\begin{equation}
H_p(f_x,f_y) = \Circ  \left( \frac{ \lambda z_2 \sqrt{f_x^2 +f_y^2}}{a} \right) 
\end{equation}
The cut-off frequency is given by:
\begin{equation}
f_{max}=\frac{a}{\lambda z_2}
\end{equation}

In digital imaging, the above mentioned optical resolution (cycles/mm) is the limit only when it is lower than digital resolution limit which is the ratio of pixel pitch and magnification. Magnification is defined by the ratio of source-sensor distance and source-object distance.
\begin{figure}
\centering
\includegraphics[width=8.25 cm]{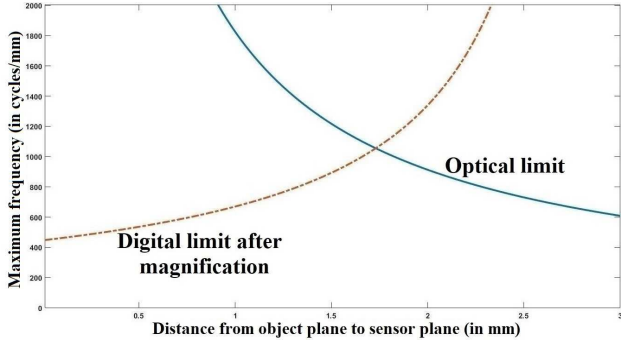}
\caption{Graph showing the variation of optical resolution and digital resolution with $ z_2$  for a fixed $z=3$ mm.}
\label{fig:false-color}
\end{figure}
Graph shown in Figure 6 tells us that the final resolution is best achieved here when the magnification is close to 2.4. This corresponds to a half-pitch resolution of $\sim$500 nm. Note that the digital limit in this figure refers to the ratio of pixel-pitch and magnification and the optical limit is given by equation 24.

Note that in the computational methods of microscopy, resolution is further improved during true the image estimation process and this step is basically affected by the noise and a priori information about the object.

\subsection{Working distance}
An interesting result in gaussian beam based lensless imaging is that the resolution is independent of the actual values of $z$ and $z_2$ but instead depends on the ratio. This can be seen for optical resolution by rearranging the equation 22.
\begin{equation}
f_{max}=\frac{z}{z_2} \frac{ \tan \left( \arcsin NA_{OF} \right)}{\lambda}
\end{equation}
where $NA_{OF} $ is the numerical aperture of optical fiber used as the source.
Digital resolution also depends on the ratio of $z$ and $z_1$ instead of their actual values.

This shows that a larger working distance can be obtained by scaling up $z$ and $z_2$ without compromising resolution with same optical fiber source. This scaling up will also lead to an increased FOV. Here in this paper, we have demonstrated $z= \sim 3.4$ mm and $z_1=z_2= \sim1.7$ mm. The limit on this scaling up is defined by the following factors: First, the physical size of image sensor i.e. aperture of gaussian beam should be smaller or equal to the image sensor dimensions. Second, the coherence length of light should be long enough to record the spectrum in the hologram to the extent of above defined bandwidth (see reference \cite{agbana2017aliasing} for detail on this). Also, for larger working distances, angular spectrum method shows accumulation of error. Various investigators have shown different solutions to reduce these errors \cite{matsushima2009band,kozacki2012computation}.

\section{Conclusions}
The presented method of lensless digital in-line holographic microscopy used photonic crystal fiber as high numerical aperture (NA) coherent light source together with angular spectrum method and  maximum-likelihood blind deconvolution algorithm. This led to an improvement in both lateral resolution and working distance as compared to previous methods of single-shot lensless imaging. Field of view is reduced by the factor of magnification but is still better than the FOV of 20X objective (at lensless magnification =2).

\section{Acknowledgments}
Sanjeev Kumar acknowledges Council of Scientific and Industrial Research (CSIR), India (File No: 09/081(1282)/2016-EMR-1) for the award of an individual senior research fellowship. All the authors acknowledge Prof. Soumen Das, Ms. Jyotsana Priyadarshani and Mr. Prasoon Awasthi; Bio-MEMS lab, IIT-Kharagpur for helping with the microscopy. All the authors thank the anonymous reviewers.

\bibliographystyle{elsarticle-num-names}
\bibliography{Sanjeev_OFT_arxiv}

\begin{thebibliography}{19}
\expandafter\ifx\csname natexlab\endcsname\relax\def\natexlab#1{#1}\fi
\providecommand{\url}[1]{\texttt{#1}}
\providecommand{\href}[2]{#2}
\providecommand{\path}[1]{#1}
\providecommand{\DOIprefix}{doi:}
\providecommand{\ArXivprefix}{arXiv:}
\providecommand{\URLprefix}{URL: }
\providecommand{\Pubmedprefix}{pmid:}
\providecommand{\doi}[1]{\href{http://dx.doi.org/#1}{\path{#1}}}
\providecommand{\Pubmed}[1]{\href{pmid:#1}{\path{#1}}}
\providecommand{\bibinfo}[2]{#2}
\ifx\xfnm\relax \def\xfnm[#1]{\unskip,\space#1}\fi
\bibitem[{Gabor et~al.(1948)}]{gabor1948new}
\bibinfo{author}{D.~Gabor}, et~al.,
\newblock \bibinfo{title}{A new microscopic principle},
\newblock \bibinfo{journal}{Nature} \bibinfo{volume}{161}
  (\bibinfo{year}{1948}) \bibinfo{pages}{777--778}.
\bibitem[{Fienup(1982)}]{fienup1982phase}
\bibinfo{author}{J.~R. Fienup},
\newblock \bibinfo{title}{Phase retrieval algorithms: a comparison},
\newblock \bibinfo{journal}{Applied optics} \bibinfo{volume}{21}
  (\bibinfo{year}{1982}) \bibinfo{pages}{2758--2769}.
\bibitem[{Latychevskaia and Fink(2007)}]{latychevskaia2007solution}
\bibinfo{author}{T.~Latychevskaia}, \bibinfo{author}{H.-W. Fink},
\newblock \bibinfo{title}{Solution to the twin image problem in holography},
\newblock \bibinfo{journal}{Physical review letters} \bibinfo{volume}{98}
  (\bibinfo{year}{2007}) \bibinfo{pages}{233901}.
\bibitem[{Maiden et~al.(2011)Maiden, Humphry, Zhang, and
  Rodenburg}]{maiden2011superresolution}
\bibinfo{author}{A.~M. Maiden}, \bibinfo{author}{M.~J. Humphry},
  \bibinfo{author}{F.~Zhang}, \bibinfo{author}{J.~M. Rodenburg},
\newblock \bibinfo{title}{Superresolution imaging via ptychography},
\newblock \bibinfo{journal}{JOSA A} \bibinfo{volume}{28} (\bibinfo{year}{2011})
  \bibinfo{pages}{604--612}.
\bibitem[{Luo et~al.(2015)Luo, Greenbaum, Zhang, and Ozcan}]{luo2015synthetic}
\bibinfo{author}{W.~Luo}, \bibinfo{author}{A.~Greenbaum},
  \bibinfo{author}{Y.~Zhang}, \bibinfo{author}{A.~Ozcan},
\newblock \bibinfo{title}{Synthetic aperture-based on-chip microscopy},
\newblock \bibinfo{journal}{Light: Science \& Applications} \bibinfo{volume}{4}
  (\bibinfo{year}{2015}) \bibinfo{pages}{e261}.
\bibitem[{Ozcan and McLeod(2016)}]{ozcan2016lensless}
\bibinfo{author}{A.~Ozcan}, \bibinfo{author}{E.~McLeod},
\newblock \bibinfo{title}{Lensless imaging and sensing},
\newblock \bibinfo{journal}{Annual review of biomedical engineering}
  \bibinfo{volume}{18} (\bibinfo{year}{2016}) \bibinfo{pages}{77--102}.
\bibitem[{Jericho and J{\"u}rgen~Kreuzer(2011)}]{jericho2011point}
\bibinfo{author}{M.~H. Jericho}, \bibinfo{author}{H.~J{\"u}rgen~Kreuzer},
\newblock \bibinfo{title}{Point source digital in-line holographic microscopy},
\newblock \bibinfo{journal}{Coherent Light Microscopy}  (\bibinfo{year}{2011})
  \bibinfo{pages}{3--30}.
\bibitem[{Takaki and Ohzu(1999)}]{takaki1999fast}
\bibinfo{author}{Y.~Takaki}, \bibinfo{author}{H.~Ohzu},
\newblock \bibinfo{title}{Fast numerical reconstruction technique for
  high-resolution hybrid holographic microscopy},
\newblock \bibinfo{journal}{Applied optics} \bibinfo{volume}{38}
  (\bibinfo{year}{1999}) \bibinfo{pages}{2204--2211}.
\bibitem[{Kanka et~al.(2009)Kanka, Riesenberg, and
  Kreuzer}]{kanka2009reconstruction}
\bibinfo{author}{M.~Kanka}, \bibinfo{author}{R.~Riesenberg},
  \bibinfo{author}{H.~Kreuzer},
\newblock \bibinfo{title}{Reconstruction of high-resolution holographic
  microscopic images},
\newblock \bibinfo{journal}{Optics letters} \bibinfo{volume}{34}
  (\bibinfo{year}{2009}) \bibinfo{pages}{1162--1164}.
\bibitem[{Kanka et~al.(2011)Kanka, Riesenberg, Petruck, and
  Graulig}]{kanka2011high}
\bibinfo{author}{M.~Kanka}, \bibinfo{author}{R.~Riesenberg},
  \bibinfo{author}{P.~Petruck}, \bibinfo{author}{C.~Graulig},
\newblock \bibinfo{title}{High resolution (na= 0.8) in lensless in-line
  holographic microscopy with glass sample carriers},
\newblock \bibinfo{journal}{Optics letters} \bibinfo{volume}{36}
  (\bibinfo{year}{2011}) \bibinfo{pages}{3651--3653}.
\bibitem[{Goodman(2005)}]{goodman2005introduction}
\bibinfo{author}{J.~W. Goodman}, \bibinfo{title}{Introduction to Fourier
  optics}, \bibinfo{publisher}{Roberts and Company Publishers},
  \bibinfo{year}{2005}.
\bibitem[{Matsushima and Shimobaba(2009)}]{matsushima2009band}
\bibinfo{author}{K.~Matsushima}, \bibinfo{author}{T.~Shimobaba},
\newblock \bibinfo{title}{Band-limited angular spectrum method for numerical
  simulation of free-space propagation in far and near fields},
\newblock \bibinfo{journal}{Optics express} \bibinfo{volume}{17}
  (\bibinfo{year}{2009}) \bibinfo{pages}{19662--19673}.
\bibitem[{Kozacki et~al.(2012)Kozacki, Falaggis, and
  Kujawinska}]{kozacki2012computation}
\bibinfo{author}{T.~Kozacki}, \bibinfo{author}{K.~Falaggis},
  \bibinfo{author}{M.~Kujawinska},
\newblock \bibinfo{title}{Computation of diffracted fields for the case of high
  numerical aperture using the angular spectrum method},
\newblock \bibinfo{journal}{Applied optics} \bibinfo{volume}{51}
  (\bibinfo{year}{2012}) \bibinfo{pages}{7080--7088}.
\bibitem[{Markham and Conchello(1999)}]{markham1999parametric}
\bibinfo{author}{J.~Markham}, \bibinfo{author}{J.-A. Conchello},
\newblock \bibinfo{title}{Parametric blind deconvolution: a robust method for
  the simultaneous estimation of image and blur},
\newblock \bibinfo{journal}{JOSA A} \bibinfo{volume}{16} (\bibinfo{year}{1999})
  \bibinfo{pages}{2377--2391}.
\bibitem[{Nikon(????)}]{Nikon}
\bibinfo{author}{Nikon}, \bibinfo{title}{Cfi60 optics}, ???? \URLprefix
  \url{https://www.nikoninstruments.com/images/stories/PDFs/cfi60_2ce-mssh-4.pdf}.
\bibitem[{Born and Wolf(2013)}]{born2013principles}
\bibinfo{author}{M.~Born}, \bibinfo{author}{E.~Wolf},
  \bibinfo{title}{Principles of optics: electromagnetic theory of propagation,
  interference and diffraction of light}, \bibinfo{publisher}{Elsevier},
  \bibinfo{year}{2013}.
\bibitem[{K{\"o}hler(1981)}]{kohler1981abbe}
\bibinfo{author}{H.~K{\"o}hler},
\newblock \bibinfo{title}{On abbe's theory of image formation in the
  microscope},
\newblock \bibinfo{journal}{Journal of Modern Optics} \bibinfo{volume}{28}
  (\bibinfo{year}{1981}) \bibinfo{pages}{1691--1701}.
\bibitem[{Horstmeyer et~al.(2016)Horstmeyer, Heintzmann, Popescu, Waller, and
  Yang}]{horstmeyer2016standardizing}
\bibinfo{author}{R.~Horstmeyer}, \bibinfo{author}{R.~Heintzmann},
  \bibinfo{author}{G.~Popescu}, \bibinfo{author}{L.~Waller},
  \bibinfo{author}{C.~Yang},
\newblock \bibinfo{title}{Standardizing the resolution claims for coherent
  microscopy},
\newblock \bibinfo{journal}{Nature Photonics} \bibinfo{volume}{10}
  (\bibinfo{year}{2016}) \bibinfo{pages}{68--71}.
\bibitem[{Agbana et~al.(2017)Agbana, Gong, Amoah, Bezzubik, Verhaegen, and
  Vdovin}]{agbana2017aliasing}
\bibinfo{author}{T.~E. Agbana}, \bibinfo{author}{H.~Gong},
  \bibinfo{author}{A.~S. Amoah}, \bibinfo{author}{V.~Bezzubik},
  \bibinfo{author}{M.~Verhaegen}, \bibinfo{author}{G.~Vdovin},
\newblock \bibinfo{title}{Aliasing, coherence, and resolution in a lensless
  holographic microscope},
\newblock \bibinfo{journal}{Optics letters} \bibinfo{volume}{42}
  (\bibinfo{year}{2017}) \bibinfo{pages}{2271--2274}.

\end{thebibliography}

\end{document}